\documentclass[epj,referee]{svjour}
\usepackage{graphics}
\begin{document}
\title{A lattice Boltzmann model with random dynamical constraints}
\author{A. Lamura\inst{1} \and S. Succi\inst{2}         
}
\institute{Istituto Applicazioni Calcolo, CNR, Sezione di Bari,
           Via Amendola 122/D, 70126 Bari, Italy\\
           \email{a.lamura@area.ba.cnr.it} \and 
           Istituto Applicazioni Calcolo, CNR,
           V.le del Policlinico 137, 00161 Roma, Italy\\
           \email{succi@iac.rm.cnr.it}}
\date{Received: date / Revised version: date}
\abstract{
In this paper we introduce a modified lattice Boltzmann model (LBM)
with the capability of mimicking a fluid system with  
dynamic heterogeneities.
The physical system is modeled as a one-dimensional   
fluid, interacting with finite-lifetime moving obstacles.    
Fluid motion is described by a lattice Boltzmann equation and
obstacles are randomly distributed
semi-permeable barriers which constrain the motion of
the fluid particles. After a lifetime delay, obstacles move to new random 
positions. 
It is found that the non-linearly coupled dynamics of the fluid and
obstacles produces heterogeneous patterns in fluid density and 
non-exponential relaxation of two-time autocorrelation function.
}
\maketitle
\section{Introduction}

Slow relaxation to local equilibrium is a hallmark of complex
system behaviour, with many examples in physics, material science, 
and biology \cite{review}.  
From a many-body point of view, the emergence of 
long-time relaxation seems to be related to the gradual confinement
of the system in lower-dimensional regions ('slow' manifolds) of
phase-space, surrounded by isolating barriers of increasing
amplitude as the temperature is decreased (or density increased) \cite{land}.
Within this picture, long-time relaxation is often associated with
the appearance of space-time heterogeneities, which can be tentatively
attributed to self-trapping effects, i.e. molecules get trapped into
'cages' formed by high-density aggregations of other groups molecules.
While the landscape picture necessarily calls for many-body investigations
(molecular dynamics, Monte Carlo \cite{mc}
and various types of lattice-'glasses' \cite{lg}, for the case
of glasses), dynamic heterogeneities can also be interpreted in terms
of mutual interactions between fluid molecules of different mobilities
\cite{species}.
If such a picture is accepted, it is then natural to attempt a
description in terms of much simpler effective single-body 
(mean-field) approaches.
This is precisely the conceptual framework of the present work.

We develop a mesoscopic model of dynamically heterogeneous fluids based
on the lattice Boltzmann equation (LBE). 
LBE is a minimal kinetic equation describing stylized pseudo-molecules 
evolving in a regular lattice according to a simple 
local dynamics including free-streaming, collisional relaxation 
and (effective) intermolecular interactions \cite{lbe}.
LBE has proven extremely successful for a variety of complex flows,
but its applicability to disordered fluids with
long-time relaxation appears more problematic
\cite{noi}.
A crucial ingredient to produce dynamic heterogeneities is (dynamic)
geometrical frustration, i.e. the introduction of constraints which reduce
the phase-space available to the fluid system.
To model these effects, the dynamics of the LBE molecules includes 
an interaction with space-time dependent obstacles hindering 
their mobility.
As we shall see, the coexistence of 'slow' (the obstacles) and 'fast' (LBE
molecules) makes the dynamics of the present model 
depart significantly from simple fluid behavior.

\section{The model}

In the present paper, we use a modified lattice Boltzmann equation.
Standard LBE with a single relaxation time \cite{bhat54}
reads as follows (time-step made unit for simplicity):
\begin{equation} 
f_i({\bf r},t) -
f_i({\bf r}-{\bf c}_i,t-1) =
-\omega \left [ f_i-f_i^e \right ] ({\bf r}-{\bf c}_i,t-1) 
\label{lbe}
\end{equation} 
where $f_i({\bf r},t) \equiv f({\bf r},{\bf v}={\bf c}_i,t)$ is a
discrete distribution function of particles
moving along the direction $i$ with discrete speed ${\bf c}_i$.
The right-hand side represents
the relaxation to a local equilibrium $f_i^e$
in a time lapse of the order of $\omega^{-1}$.

The equilibrium distribution functions $f_{i}^{e}$
are expressed as series expansions of the Maxwellian
distribution function, up to second order with respect to the
local velocity ${\bf u}$ \cite{qian}:
\begin{equation}
f_i^e = \rho w_i [1+\frac{{\bf u} \cdot {\bf c}_i}{c_s^2}
+\frac{{\bf u}{\bf u} \cdot ({\bf c}_i {\bf c}_i-c_s^2 I)}{2 c_s^4}]
\label{fequil}
\end{equation}
The local density $\rho=\rho({\bf r},t)$, as well as 
the local velocity ${\bf u}$ which enter eq.~(\ref{fequil}), are calculated
from the distribution functions as follows:
\begin{eqnarray}
\rho & = & \sum_{i} f_{i} , \\
{\bf u} & = & \frac{1}{\rho}\sum_{i} f_{i} {\bf c}_{i} .
\end{eqnarray}
$w_i$ is a set of weights normalized to unity,
$c_s$ the lattice sound speed, 
and finally, $I$ stands for the unit tensor.
In the following, we shall refer to
a one-dimensional lattice of size $L$ with $c_i=-1,0+1$ and
$c_s=1/\sqrt{3}$.

\begin{figure}
\begin{center}
\resizebox{0.33\textwidth}{!}{%
  \includegraphics{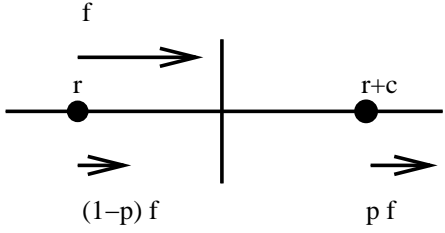}
}
\end{center}
\caption{Schematic representation of dynamical role of obstacles
during propagation of distribution functions.
}
\label{fig_coll}
\end{figure}

We generalize the above eq.~(\ref{lbe}) as follows
\begin{eqnarray} 
&&f_i(r,t)-\left [ 1-p(r+\frac{ c_i}{2},t-1)
\right ] 
f_i(r,t-1) = \nonumber \\
&&p(r-\frac{c_i}{2},t-1) \Big [ f_i(r-c_i,t-1)\nonumber \\
&&-\omega \left [ f_i-f_i^e \right ] (r-c_i,t-1) \Big ]
\label{lbm}
\end{eqnarray} 
The variables $p(r \pm \frac{c_i}{2},t)$ 
live on the lattice links and vary in space and time
(non-quenched disorder \cite{disord}). We choose them in the form
of binary fields taking only the values
$p=1$ and $p=p_t$ with $0 < p_t < 1$. 
Links with $p = p_t$ act as semi-permeable obstacles
which control the propagation rate
between neighboring sites in a way which preserves the total
density conservation.
Figure~\ref{fig_coll} illustrates in a pictorial way the role
of obstacles.
Obstacles are characterized by their {\it permeability}, $p_t$, 
and {\it concentration}, $c=O/L$, $O$ being the number of obstacles,
which is kept fixed in time.

The dynamics of the obstacles is the following. 
At time $t=0$, the number of obstacles, $O$, is chosen
and their positions are randomly selected.
At subsequent times $t_n$, $n=1,2,..$, where 
\begin{eqnarray}
t_1 &=& \tau_0 \nonumber \\
t_{n+1} &=& t_n + \tau(t_n) 
\label{ts} \\
\tau(t_n) &=& int \Big(\tau_0 e^{\frac{1}{2\rho_0}  
\big(\frac{2}{|m(t_n)-2|}-1\big)}\Big) \nonumber
\end{eqnarray}
the obstacles are shifted rightward to new randomly chosen positions
$r_j(t_{n+1})=r_j(t_n)+r'_j(t_{n+1})$, $j=1,...,O$, 
where $r'_j(t_{n+1})=int(s)+1$, $s$ being 
a random number drawn from an uniform distribution in the range
$[\frac{d}{2}, \frac{3d}{2}]$ and 
$d = \frac{L}{O}$ is the mean inter-obstacle distance. 
In the above, $int(s)$ denotes the integer part
of the variable $s$. 
The motion of each obstacle can be seen as a continuous time random walk
\cite{metzler}.
In eqs.~(\ref{ts}) $\tau_0$ is the first lifetime
and the quantity 
\begin{equation}
m(t_n)=(\rho_{max}(t_n)-\rho_{min}(t_n))/\rho_0 
\end{equation}
is the relevant order parameter.
Here, $\rho_{max}(t_n)$ and $\rho_{min}(t_n)$ are 
the maximum and minimum values of fluid 
density at time $t_n$, respectively, and
$\rho_0$ is the average fluid density.
Non zero values of $m$ are the prime indicators of departure 
from ideal fluid behavior (see below). 
The quantity $\tau_0$  has to be generally much larger than the time
scale of fluid motion so as to characterize obstacles as the slowly
moving species of particles. 
The dependence of $\tau$ on the order parameter $m$
is intended to slow down the dynamics of obstacles in the presence of density
contrasts (non zero values of $m$). Indeed, 
in the expression of $\tau(t_n)$ in (\ref{ts}) it appears a singularity at
$m=2$.
The reason is the following.
We assumed that, on the average,  $\rho_{max} - \rho_0 \simeq \rho_0 - 
\rho_{min}$ 
and later 
numerically verified this assumption to be correct.
Since it must be $\rho_{min} \ge 0$, this means that 
in the limit $\rho_{min} \rightarrow 0$ 
($\rho_{max} \rightarrow 2 \rho_0$), namely  
$m \rightarrow 2$, the lifetime of the obstacles must
diverge, $\tau(t_n) \rightarrow +\infty$.

In moving obstacles we have taken into account the periodicity
of the lattice. When a link is already occupied, the moving obstacle is
shifted to the nearest neighbor link.
The fact that obstacles move rightwards has no effect on the overall fluid
dynamics, which depends only on the position of obstacles, 
and not on their motion. 
The equation of motion of the obstacles is:
\begin{equation} 
\label{PMOTION}
p(r+r',t_{n+1}) = 
p(r,t_n), \;\;\;\; n=1,2,... 
\label{q}
\end{equation}
where the distance $r'$ has been previously defined.
The eq.~(\ref{q}) is coupled
to eq.~(\ref{lbm}) via $\tau(t_n)$, which depends on density through the 
order parameter $m=m(\rho)$.

Before proceeding further, a few distinguished limits in 
the phase-plane $p_t-c$ are worth a brief comment:

- {\it Fluid limit}: Along the lines $c=0$ (obstacle-free) 
  and $p_t=1$ (transparent obstacles) the system 
  behaves like a standard LBE fluid.

- {\it Slow-fluid limit}: Along the line $c=1$, the system behaves like a
  standard LB fluid, only with a rescaled speed
  $p_t u$. Thus, as $p_t \rightarrow 0$, this slow fluid 
  goes smoothly into the 'frozen' limit represented 
  by the corner $(p_t=0,c=1)$, where the system is frozen to its
  initial configuration.

- {\it Arrest limit}: Along the line $p_t=0$ 
  the system is liable to develop density accumulations
  in a finite time at the locations where the obstacles reside.
  The actual onset of these density pile-ups depends on the
  lifetime of the traps, as well as on their concentration.
  It is in fact clear that dilute obstacles with
  short lifetimes permit the system to release density 
  pile-ups, thereby avoiding strong density accumulations. 
  In actual practice, an 'arrest line' of the form
  $c_A=c_A(p_t,\tau_0)$ marks the upper value of the concentration
  leading to configurations without accumulation. 
  Intuitively, $c_A$ is an increasing function of $p_t$ and $\tau_0$.

The macroscopic limit of the present model remains an open question
mainly because the obstacle populations is generally smooth, so
that a standard Chapman-Enskog analysis is difficult to apply.
Here we only present a few remarks 
which apply to the limit $p_t \rightarrow 1$.
To leading order in the parameter $1-p_t$,
the main features of density profiles can be explained in terms of the 
following modified continuity equation:
\begin{equation}
\partial_t \rho + \partial_{\alpha} (p \rho u)=0 .
\label{cont}
\end{equation}
By writing the above as:
\begin{equation}
\partial_t \rho + \partial_{\alpha} (\rho u) = 
- \rho u \partial_{\alpha} p - (p-1) \partial_{\alpha} (\rho u)
\end{equation}
we recognize two extra compressibility terms on the right hand side.
The limit $p=1$ annihilates both extra compressibility contributions,
as it must be, since this is the standard LBE situation.
The case $p=p_t$ everywhere ($c=1$), leads again to a standard LBE, only with 
a rescaled speed $u \rightarrow p_t u$, hence no effective
extra-compressibility.
Genuinely extra-compressibility is therefore
associated to spatial changes of the permeability field $p(r,t)$:
$\partial_{\alpha} p \ne 0$ (this derivative must be intended in the sense
of distributions, since $p(x,t)$ is generally not smooth).

\section{Numerical results}

We performed a series of numerical simulations on a lattice
with $L=1024$ grid-points and $\omega=1.5$.
We found that results are not dependent on $\omega$, which
was varied in the range $[1,1.7]$, corresponding to a 
local collisional relaxation timescale $\tau_c \sim 1/\omega$ 
approximately in the range $[0.58,1]$ in time step units.
The parameter $\tau_0$ was fixed at $\tau_0=50$. 
The initial condition is $\rho(r)=\rho_0 \; (1+\xi(r))$
where $\xi(r)$ is a random perturbation uniformly
distributed in the range $[-0.1:0.1]$ and $\rho_0=1$.
\begin{figure*}[t]
\begin{center}
\resizebox{0.66\textwidth}{!}{%
    \includegraphics{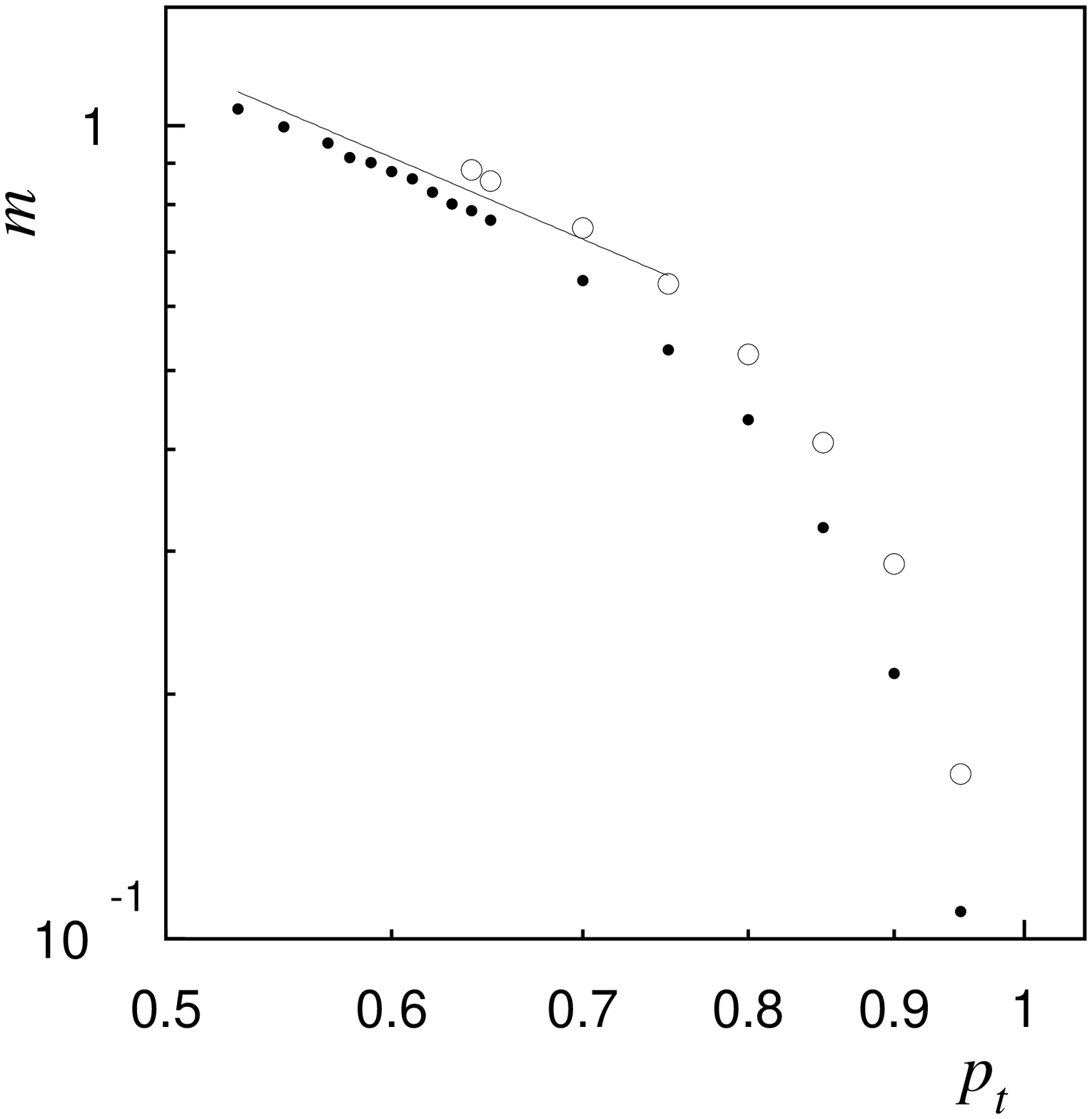}
\hskip 5.cm
    \includegraphics{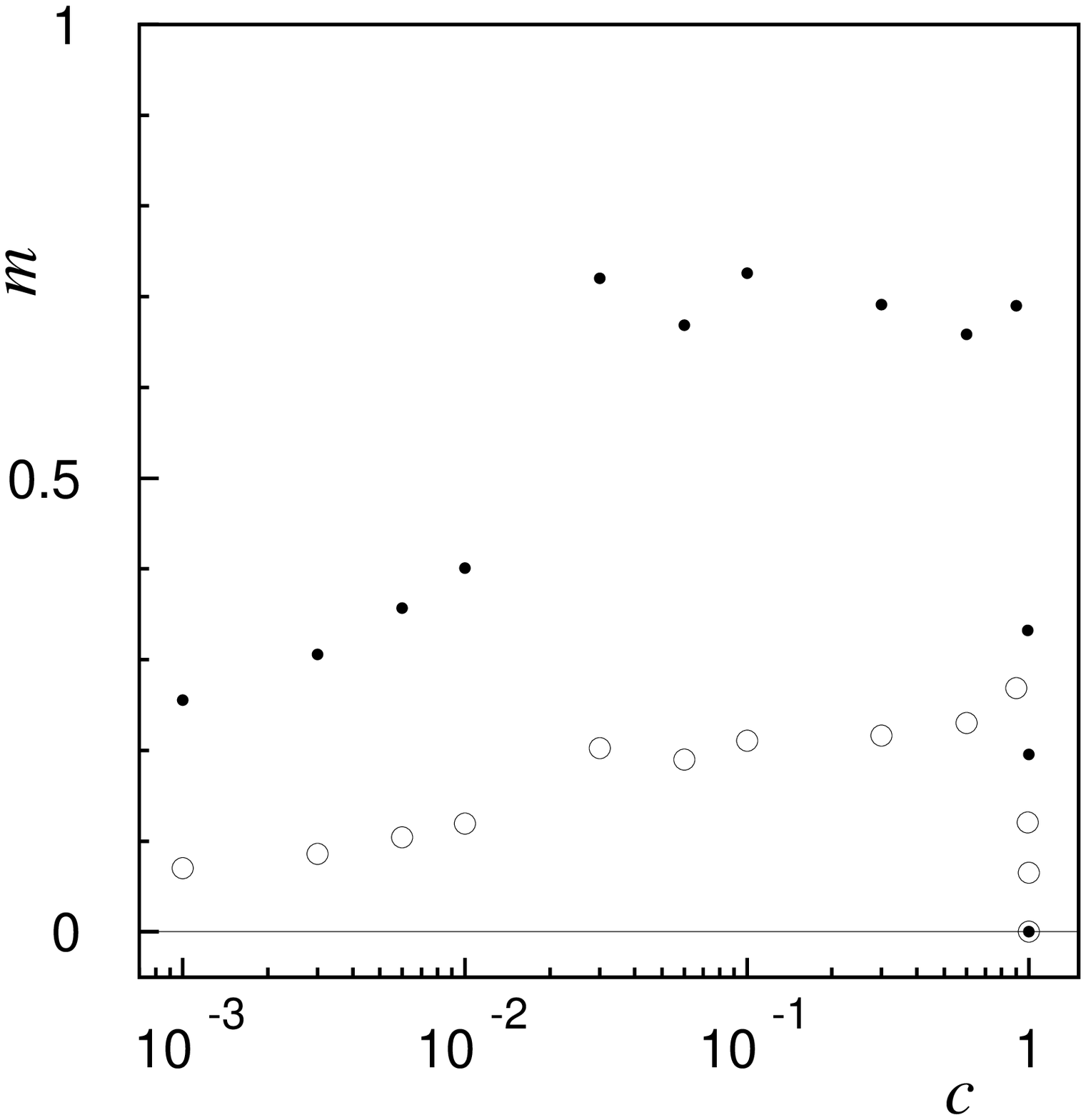}
}
\end{center}
\caption{Left panel: The order parameter $m$ as a function of $p_t$
for different concentrations $c = 0.5$ ($\bullet$), 
$0.78$ ($\circ$). The straight line has slope $-3/2$.
Right panel: The order parameter $m$ as a function of $c$
for different values of $p_t = 0.7 (\bullet), 
0.9 (\circ)$.
}
\label{fig_ord}
\end{figure*}

As a first task, we determine the phase-diagram 
of our model in the $p_t-c$ plane.
We spanned the $p_t-c$ plane and found that 
departures from ideal fluid behavior ($m \neq 0$) are observed
for every value of $(p_t,c)$ with $p_{min}<p_t<1$ and $0<c<1$. 
The limiting value (arrest value) $p_{min}$ indicates the
lowest permeability, below which the simulation is disrupted
by excessive density pile-up. 

As expected, the arrest value $p_{min}$ decreases
with decreasing $c$: We found $p_{min} \simeq 0.52$ for $c=0.5$ and 
$p_{min} \simeq 0.61$ for $c=0.78$.
The behaviors of $m$ as a function of $p_t$ and $c$ 
are shown in fig.~\ref{fig_ord}.
In the vicinity of $p_{min}$, we observe a power-law behavior
$m \sim (p_t-p_{min})^{-a}$ with $a \sim 3/2$.
Away from $p_{min}$, $m$ grows like $(p_{max}(c)-p_t)^b$, with
$b \sim 1/2$, which is essentially a mean field theory exponent,
and $p_{max}(c) \rightarrow 1$ as $c \rightarrow 1$. 
The dependence of the order parameter on $c$ at fixed $p_t$ 
is non-monotonic, with a kink around $c \sim 0.02$ and
a sharp collapse towards the 'slow-fluid' limit, $c=1$, 
which appears to be reached in a highly non-perturbative way.
To be noted that even a single obstacle $c=1/L \sim 0.001$ is sufficient to
generate sizeable non-zero values of the order parameter $m$.
In summary, these results indicate that fluid behavior
is recovered smoothly, but shows very small resilience towards 
non-zero values of the obstacle concentration $c$
and impermeability $1-p_t$. 

\begin{figure}
\begin{center}
\resizebox{0.33\textwidth}{!}{%
  \includegraphics{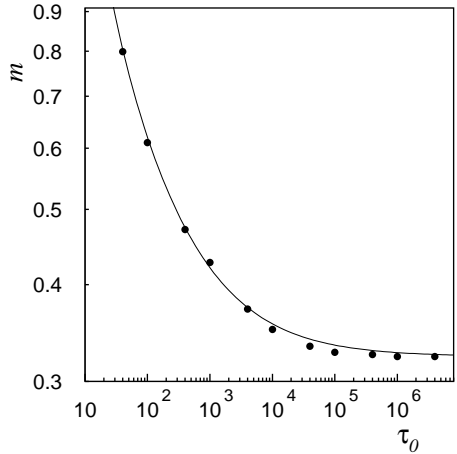}
}
\end{center}
\caption{The order parameter $m$ as a function of $\tau_0$
for $c=0.78$ and $p_t=0.7$. The full line is a guide to the eye and
represents eq.~(\ref{mmm}).
}
\label{fig_tau}
\end{figure}

We also considered the dependence of the order parameter $m$ on
the lifetime $\tau_0$. Figure~\ref{fig_tau} shows the 
behavior of $m$ as a function of the parameter
$\tau_0$ in the case with $c=0.78$, $p_t=0.7$.
It appears that $m$ decays with increasing $\tau_0$ to reach a constant value
$m(\infty)$.
The reason of this behavior is the following. In the limit of static
disorder ($\tau_0 \rightarrow \infty$) density profiles are flat
among obstacles and show sharp contrasts across them. 
Density accumulations are prevented in the long time limit 
by the flow which flattens density among obstacles. 
When reducing $\tau_0$ it is more difficult for the flow to 
smooth density profiles and the motion of obstacles produces larger
density contrasts in the system as obstacles are shifted
thereby causing $m$ to increase.
In the limit of rapid motion of obstacles ($\tau_0 \rightarrow 0$)
it happens that density contrasts are so steep that $m$ diverges.   
Further simulations with $p_t$ in the range $[0.7,0.9]$ indicate that
that $m(\infty)$ is close to the value $(1-p_t)/((1+p_t)/2)$, the
latter being an estimate of the density variation across a single obstacle
based on the continuity equation (\ref{cont}).
It is interesting to observe that the dynamic component
of the order parameter $m$ obeys a scaling law of the form
\begin{equation}
\frac{m(\tau_0)-m(\infty)}{m(\infty)}=
\frac{a}{(\tau_0-\tau_{0_{crit}})^{1/2}}
\label{mmm}
\end{equation}
again with a mean-field like exponent $1/2$, $a$ and $\tau_{0_{crit}}$ being 
two parameters which depend on $p_t$.

We next turn to the analysis of the dynamics of the model,
notably through a study of the time and space correlation
functions. To this purpose, we choose $c=0.78$, $p_t=0.7$ and $\tau_0=50$
in order to analyse a region of the phase diagram with
clear departures from ideal
fluid behavior ($m \simeq 0.75$, see figs.~\ref{fig_ord}-\ref{fig_tau}).

\begin{figure}
\begin{center}
\resizebox{0.33\textwidth}{!}{%
  \includegraphics{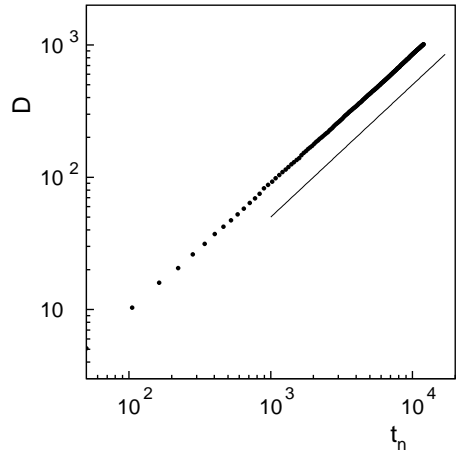}
}
\end{center}
\caption{The distance $D$ travelled by obstacles as a function of times $t_n$.
The straight line has slope $1$.
}
\label{fig_dist}
\end{figure}

We verified that the resulting motion of obstacles
can be described in terms of a directed random walk. Indeed,
we have computed the time evolution of the distance $D$ travelled by
obstacles defined as 
\begin{equation}
D(t_n)=\frac{\sum_{j=1}^{0} r_j(t_n)-r_j(0)}{O}.
\end{equation}
The quantity $D$ is plotted
in fig.~\ref{fig_dist} as a function of times $t_n$. 
\begin{figure}
\begin{center}
\resizebox{0.33\textwidth}{!}{%
  \includegraphics{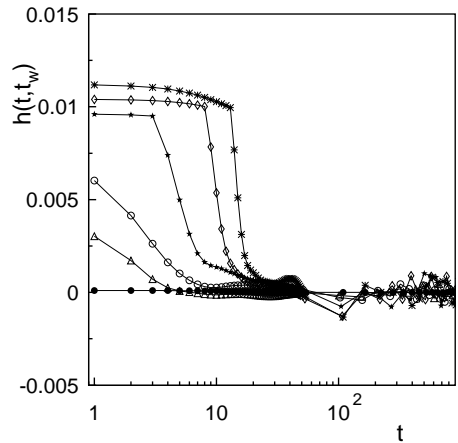}
}
\end{center}
\caption{The density autocorrelation function $h(t,t_w)$ as a function of $t$
for different waiting times $t_w$ $= 1 (\triangle)$, $52 (\circ)$, 
$104 (\star)$, 
$157 (\diamond)$, $211 (\ast)$. $h(t,t_w)$ is shown also in the case without 
obstacles for $t_w = 211 (\bullet)$.
}
\label{fig_auto}
\end{figure}
From this figure 
a power law $D \sim t$ is well visible. The exponent $1$, characterizing
the time behavior of $D$, is consistent with 
the fact that the motion of obstacles can be seen as a directed random walk 
for which the exponent is known to be $1$.
This result confirms that the obstacle motion 
is independent on fluid coupling and on obstacle shifting.

We focused our attention on the two-time density 
autocorrelation function defined as
\begin{equation}
h(t,t_w)= \frac{<\delta \rho(r,t_w) \delta \rho(r,t_w+t)>_r}
{<\rho(r,t_w)^2>_r}
\label{auto}
\end{equation}
where $\delta \rho(r)=\rho(r)-\rho_0$ is the density fluctuation around its
spatial average, 
$<...>_r$ denotes sum over the whole system and $t>t_w$.
Multiplying two configurations (one at time $t=t_w$,
the other at $t_w+t$) together site by site is sufficient to 
produce a satisfactory average 
of $h(t,t_w)$ because, due to the large number of sites ($L=1024$), the
above procedure is equivalent to obtain  $h(t,t_w)$
from an ensemble average of the system \cite{vias}.

The function $h(t,t_w)$ is not normalized to unity at $t=0$ due to the
definition (eq.~(\ref{auto})). This definition is common for
correlations of fluctuations ($\delta \rho$) since
small values of fluctuations at the denominator can give
diverging contributions to the correlation function itself.
The plot for different waiting times is presented in fig.~\ref{fig_auto}.
From this figure, evidence of non-exponential
relaxation is clearly seen, hinting at the presence of
long-lived dynamical structures.
Also, the dependence on the waiting time $t_w$ indicates the
existence of an underlying non-equilibrium, non-stationary
process. In the case without obstacles 
(standard LBE with $p=1$ in eq.~(\ref{lbm})) $h(t,t_w)$ is approximately
zero (see fig.~\ref{fig_auto}), as it should be.
\begin{figure*}
\resizebox{0.99\textwidth}{!}{%
    \includegraphics{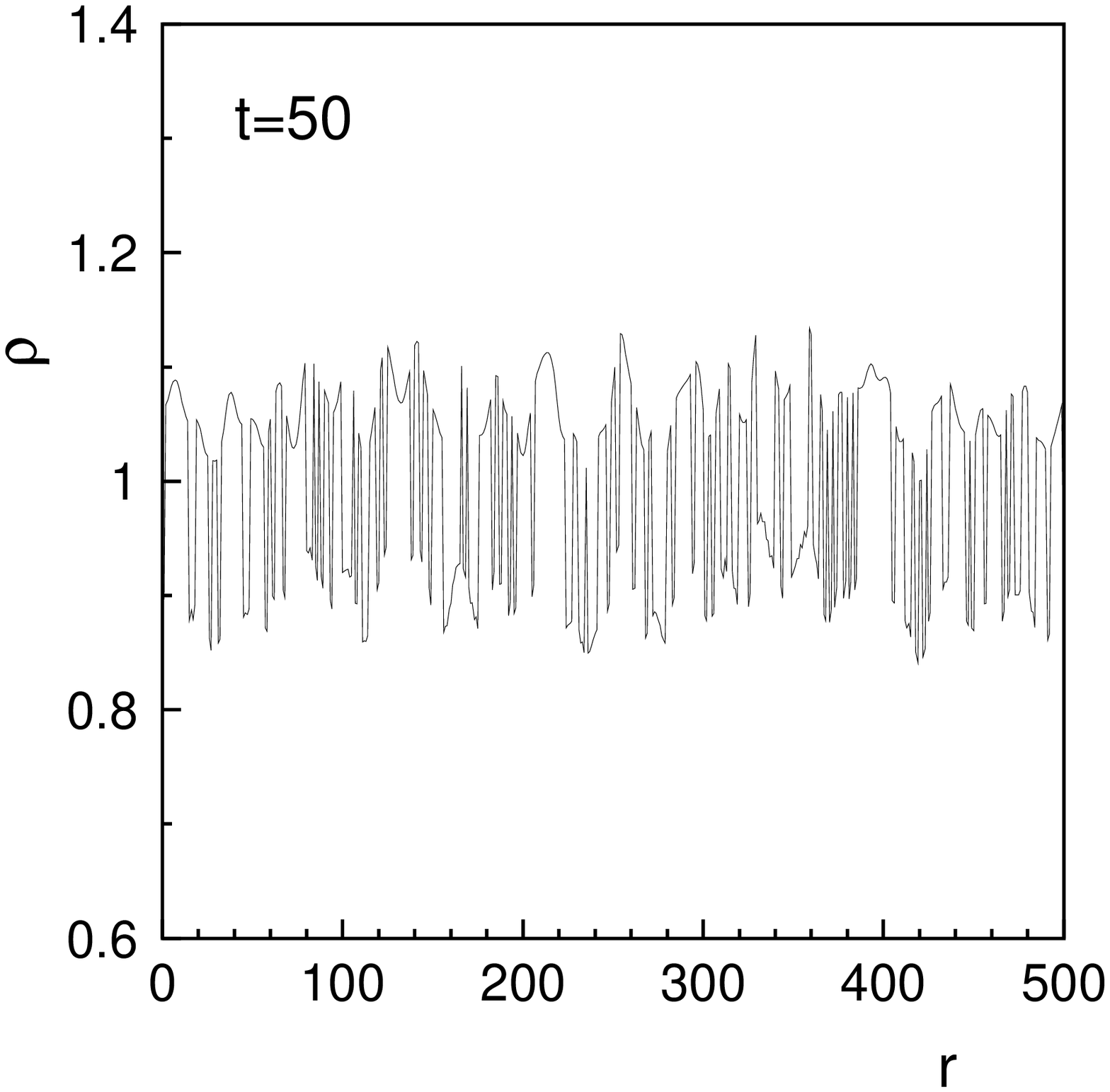}
    \includegraphics{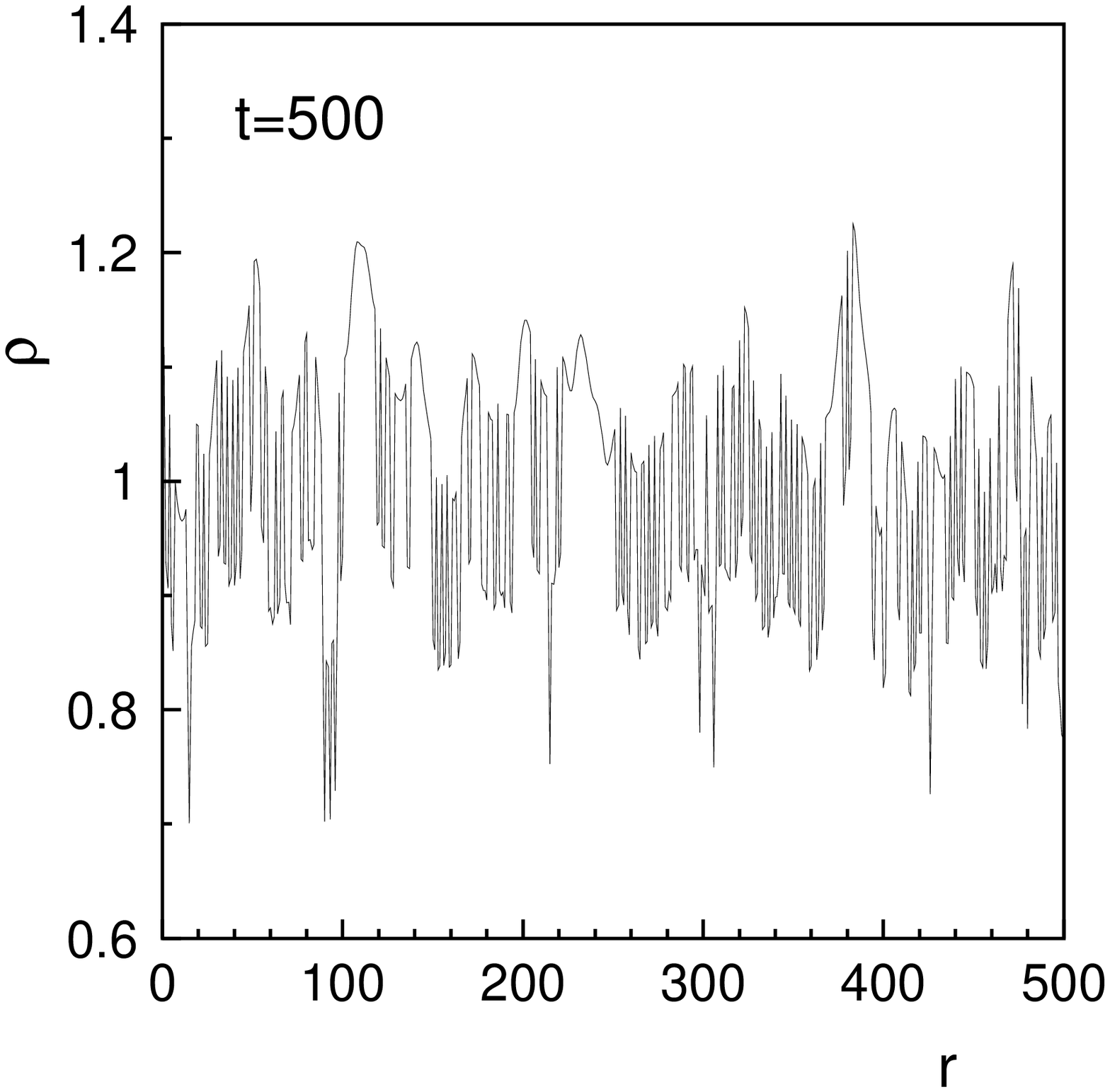}
    \includegraphics{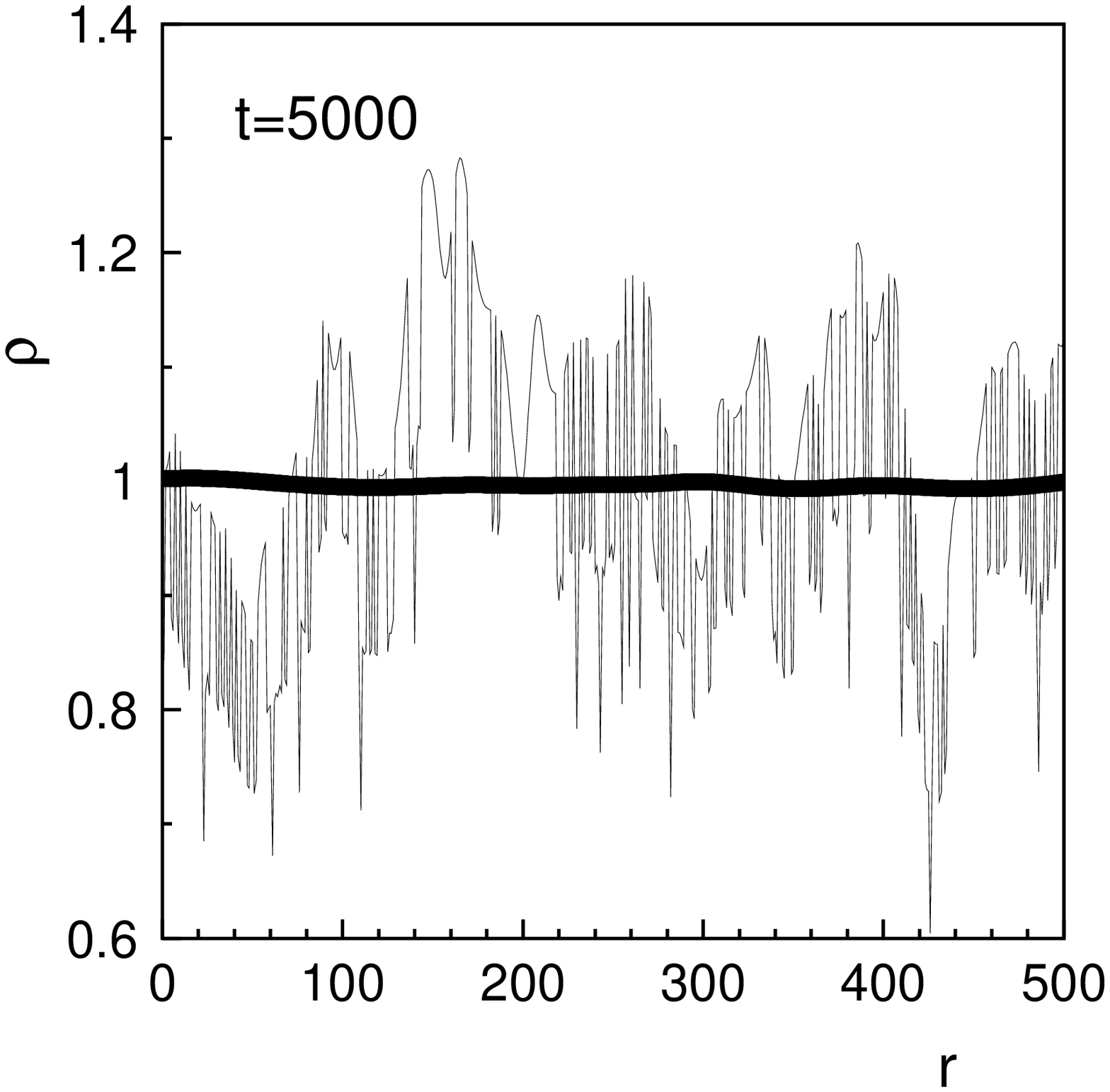}
}
\caption{Density profiles at different times on a portion of the lattice.
At time $t=5000$ the profile is shown also 
in the case without obstacles (thick line).
}
\label{fig_dens}
\end{figure*} 

\begin{figure*}
\resizebox{0.99\textwidth}{!}{%
    \includegraphics{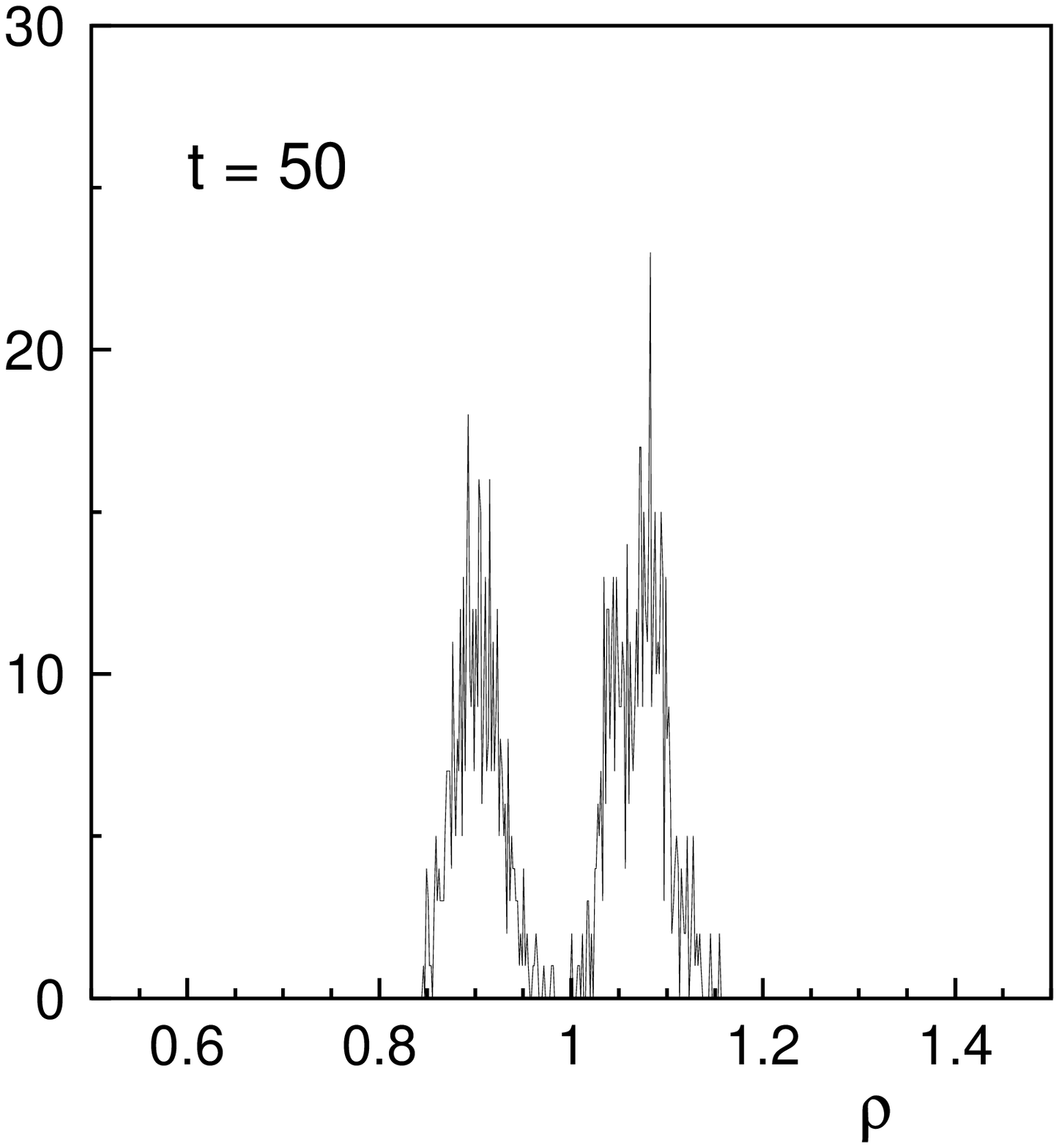}
\hskip 0.5cm
    \includegraphics{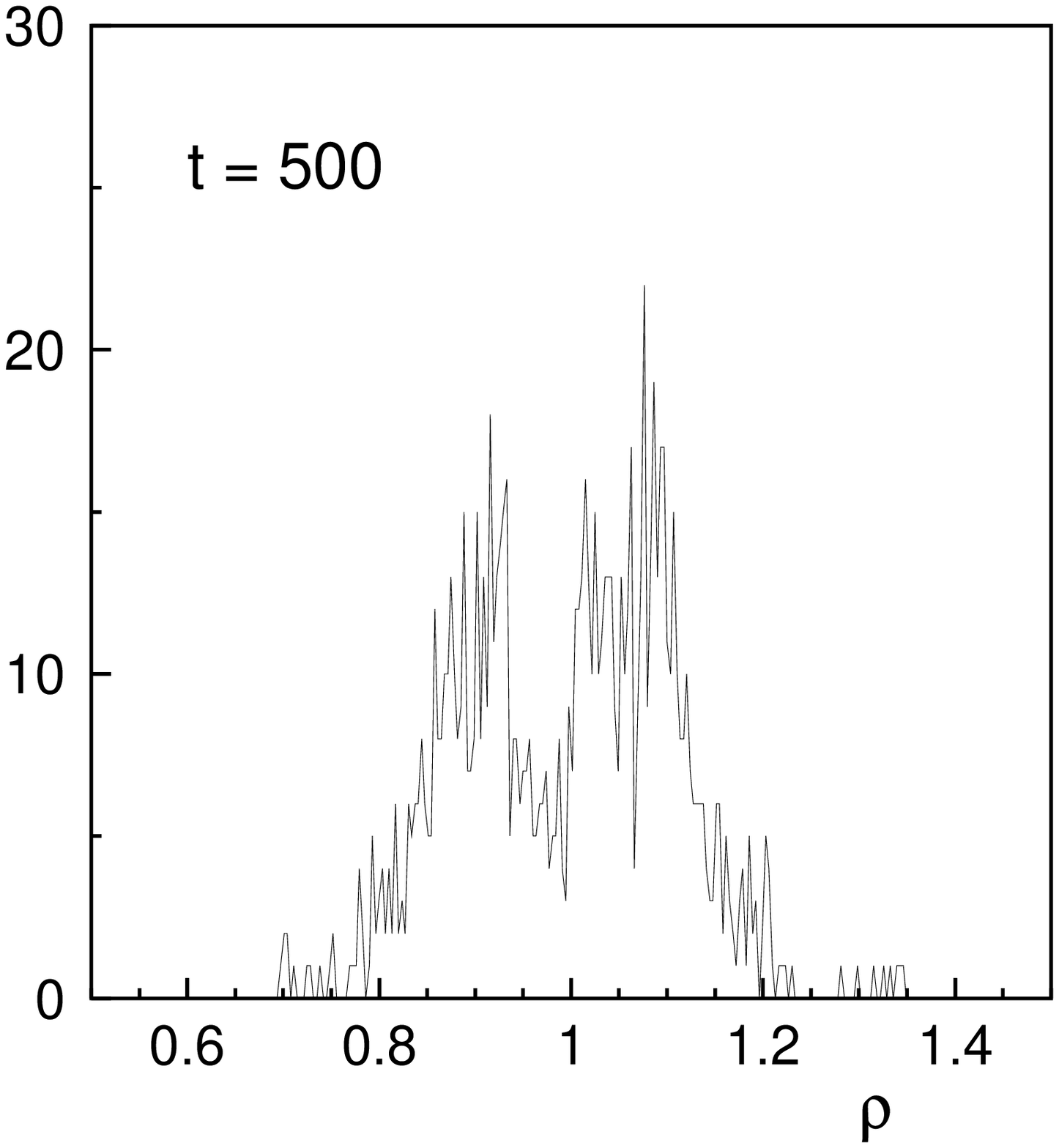}
\hskip 0.5cm
    \includegraphics{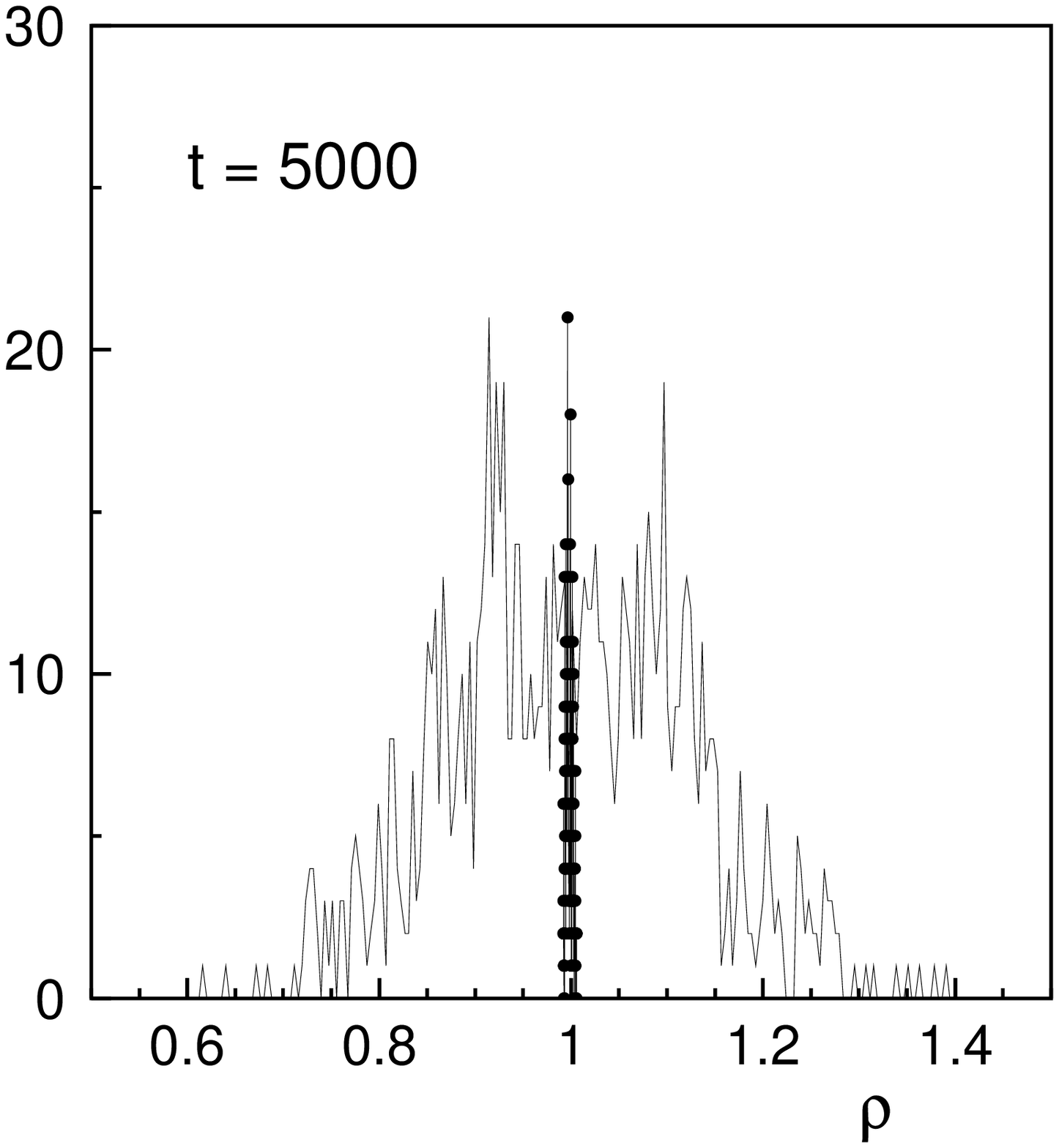}
}
\caption{The probability distribution functions (PDF) of $\rho$
at different times. At time $t=5000$ the PDF is shown also in the case
without obstacles ($-\!\!\bullet\!\!-$).
}
\label{fig_pdf}
\end{figure*}

Plots of density are shown in fig.~\ref{fig_dens} 
in the range $1-500$ for a better view.
The qualitative behavior is the same over the whole lattice.
The corresponding probability 
distribution functions (PDF) are shown in fig.~\ref{fig_pdf}.
In the initial stage the PDF quickly forgets the
initial distribution and develops two peaks
around the initial minimum and maximum densities $0.9$
and $1.1$ respectively.
The apparent phase separation at time $t=50$ is just
due to the effect of obstacles which cause density to accumulate
on some lattice sites.  
Indeed, as soon as the obstacles are released
this effect tends to fade away.
As time unfolds, these peaks 'diffuse' with the twofold
result of filling up the gap at $\rho=\rho_0$ and populating
both high and low density tails. 
It is interesting to note that the PDF generated
in these processes are {\it not} Gaussian, 
hinting at the presence of (dynamic) heterogeneities in the system. 
In the case without obstacles density $\rho$ gets constant at value
$\rho_0$ (see fig.~\ref{fig_dens} at time $t=5000$) giving a single peak
at $\rho=\rho_0$ in the PDF (see fig.~\ref{fig_pdf} at time $t=5000$).

\begin{figure*}
\resizebox{0.99\textwidth}{!}{%
    \includegraphics{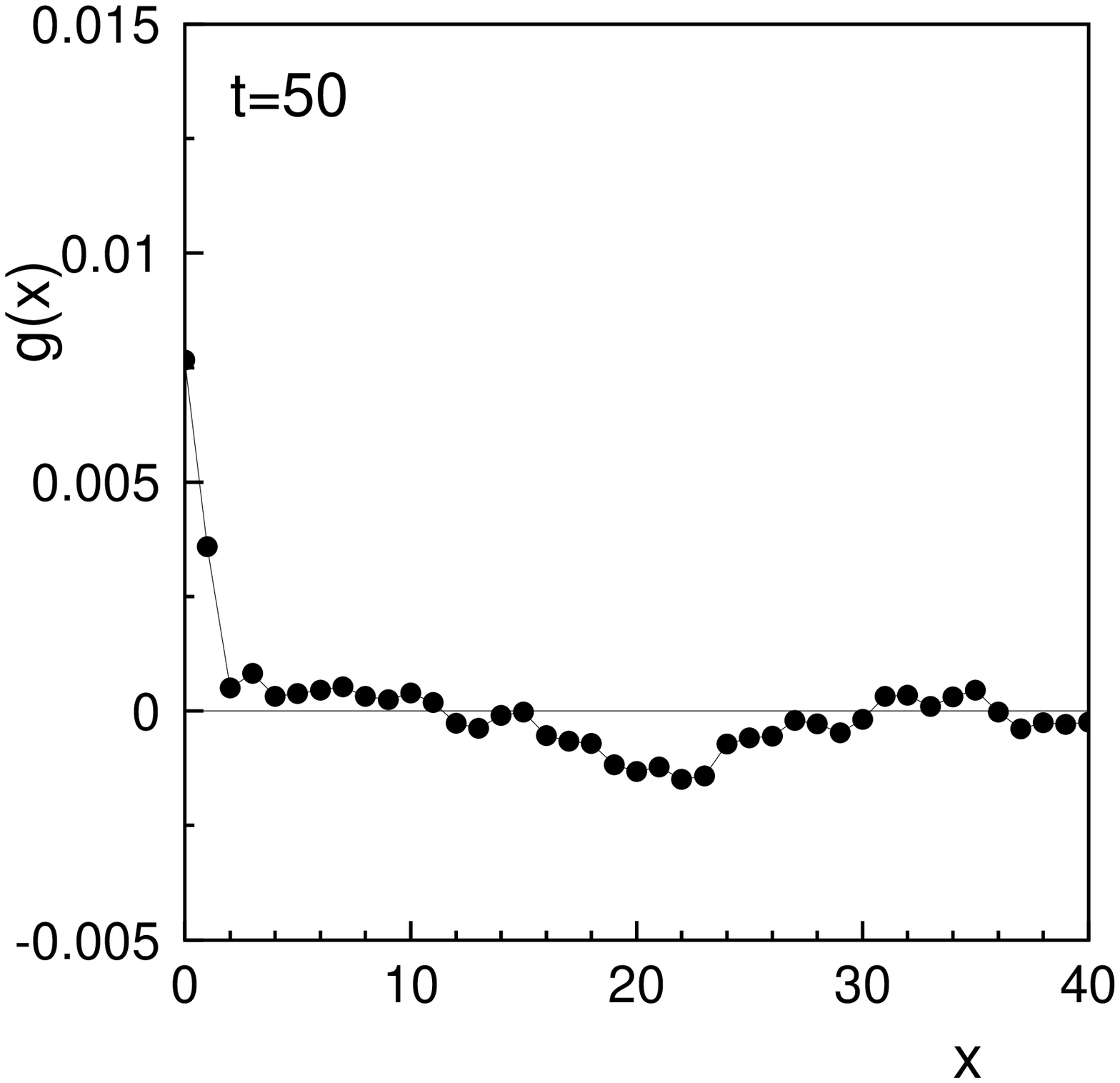}
\hskip 0.5cm
    \includegraphics{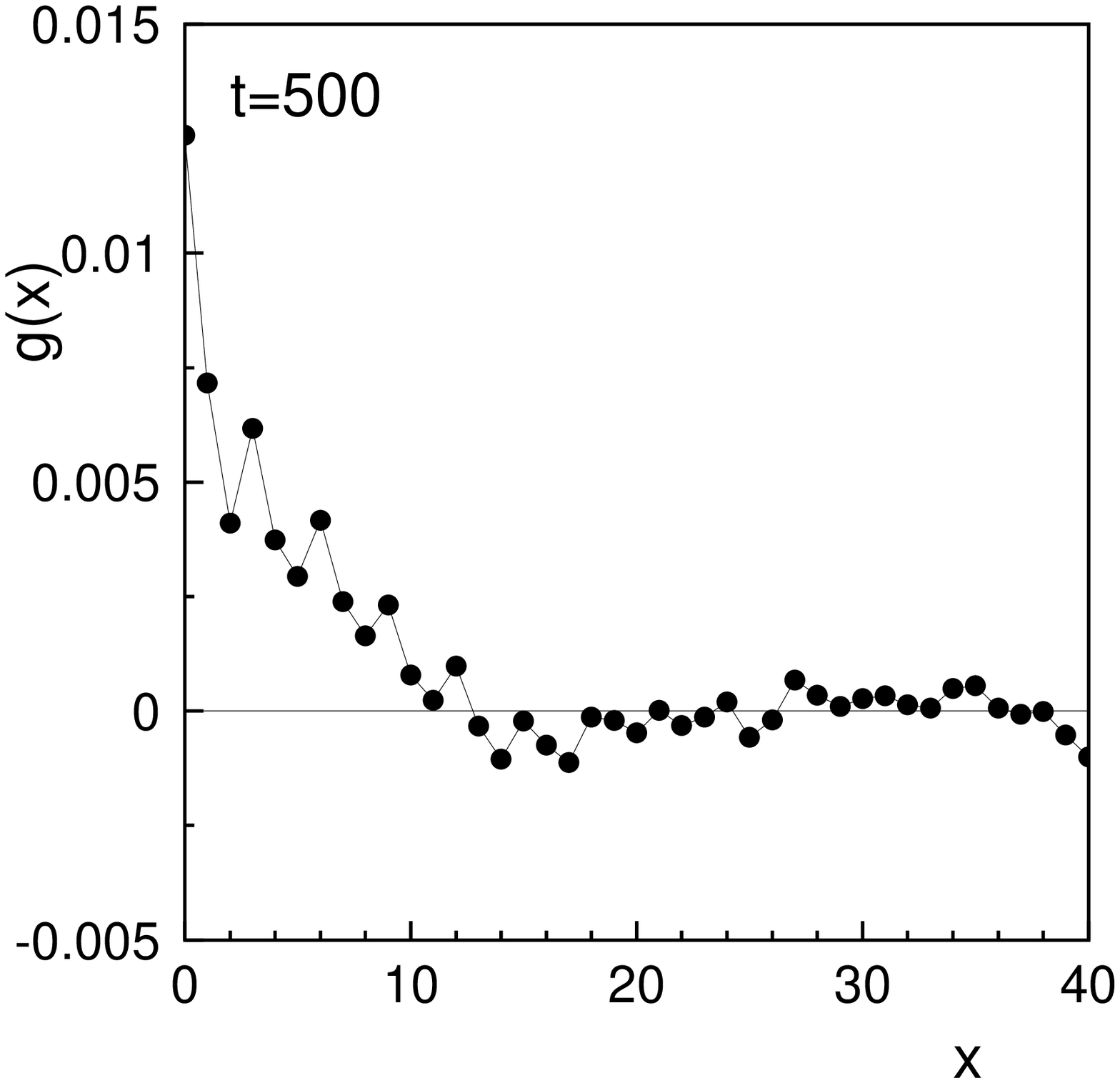}
\hskip 0.5cm
    \includegraphics{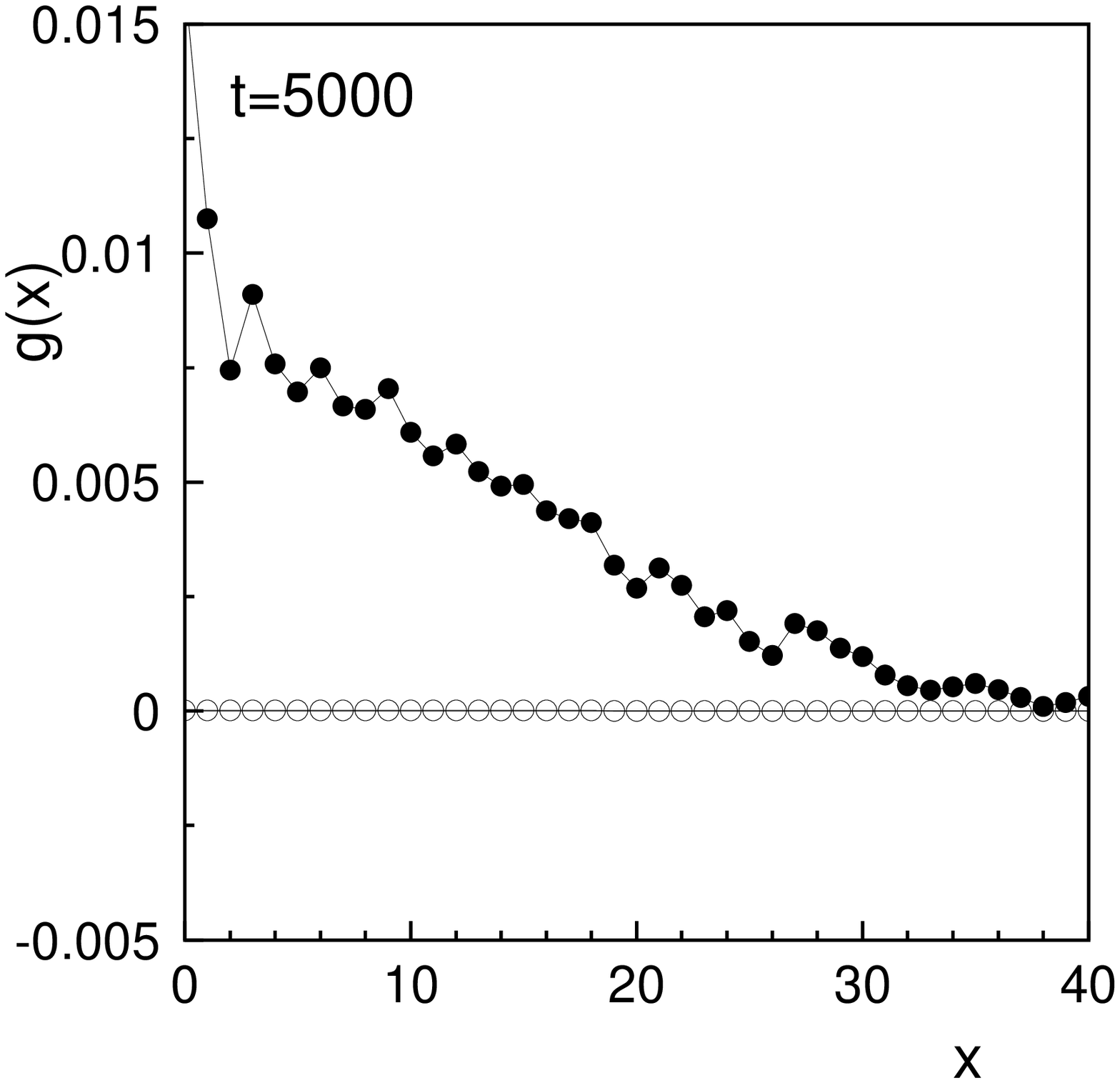}
}
\caption{The density correlation function $g(x)$ as a function of $x$
for different times. At time $t=5000$ $g(x)$ is shown also in the case
without obstacles ($\circ$). 
}
\label{fig_corr}
\end{figure*}

As a further observable, in fig.~\ref{fig_corr} we plot the 
density correlation function:
\begin{equation}
g(x)=\frac{<\delta \rho(r+x) \delta \rho(r)>_r}
{<\rho(r)^2>_r}.
\end{equation}
From these pictures, the onset of long-range order
is well appreciated. 
In the case without obstacles $g(x)$ is approximately
zero (see fig.~\ref{fig_corr} at time $t=5000$).

\begin{figure*}
\begin{center}
\resizebox{0.66\textwidth}{!}{%
    \includegraphics{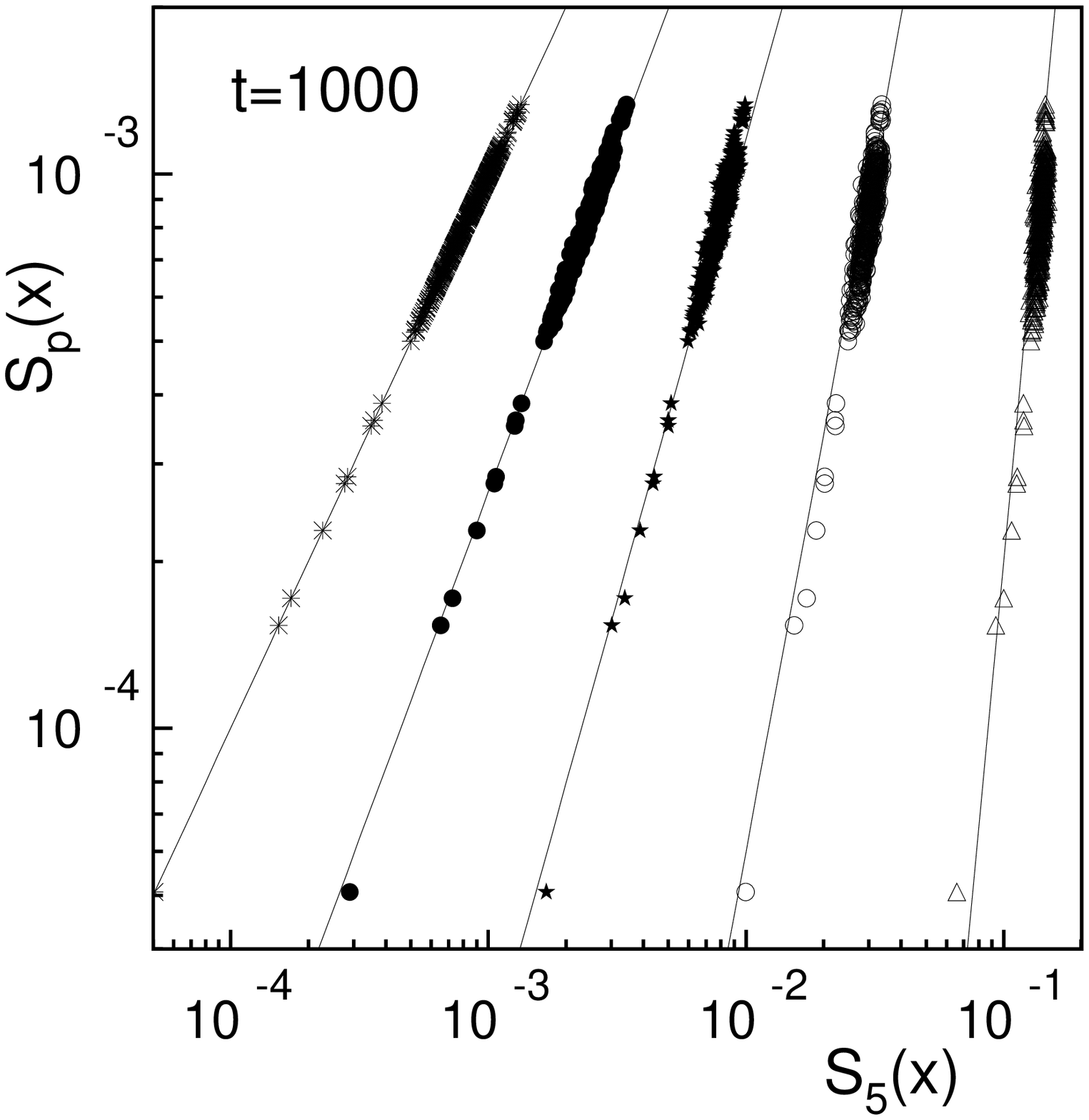}
\hskip 5.cm
    \includegraphics{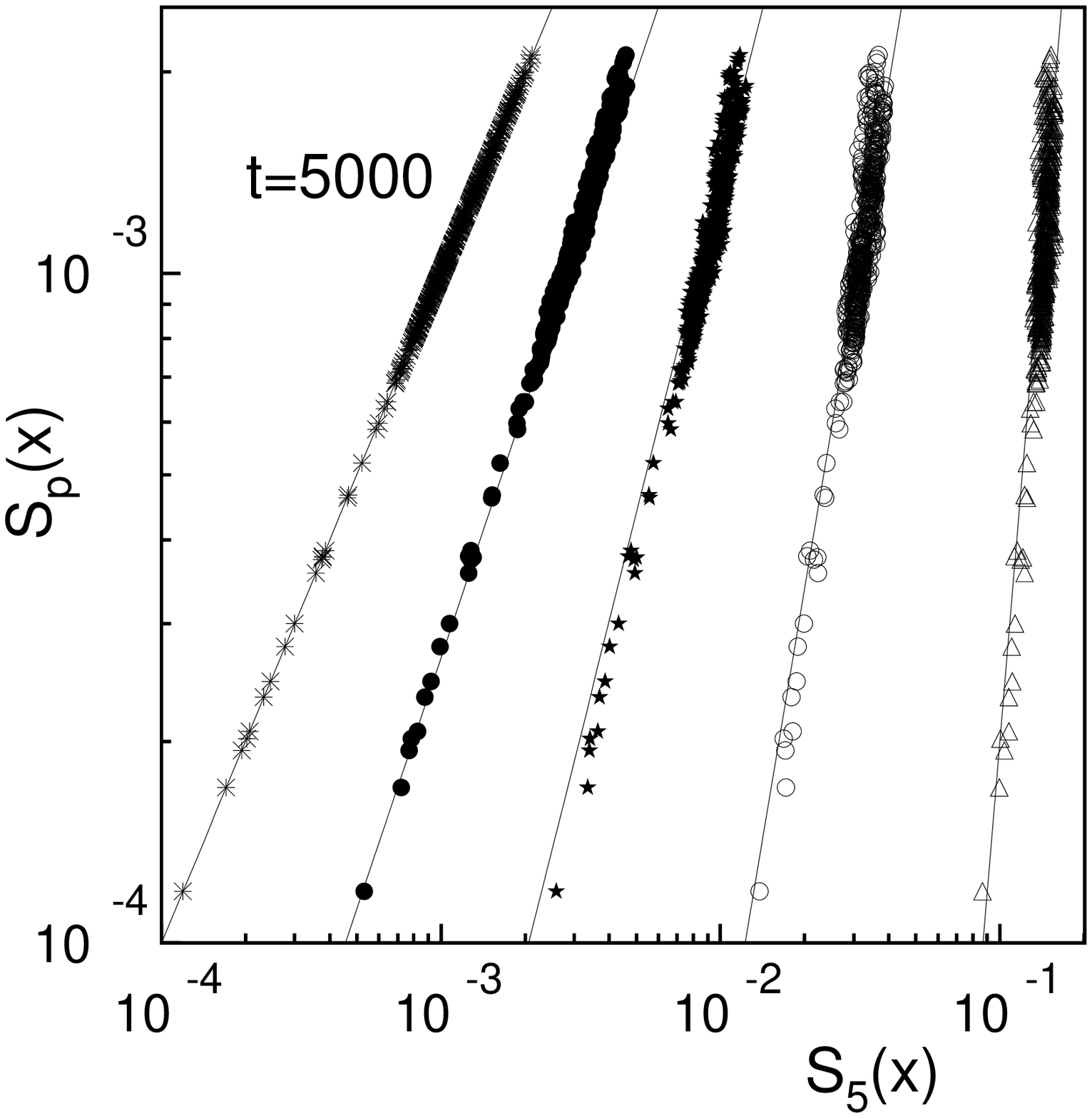}
}
\end{center}
\caption{The structure factors $S_p(x,t)$ as a function of $S_5(x,t)$ for
different values of $p= 1 (\triangle)$, $2 (\circ)$, $3 (\star)$, 
$4 (\bullet)$, 
$5 (\ast)$
at two times. The straight line have slopes $5/p$, $p=1,2,..,5$ from right 
to left.
}
\label{fig_fratt}
\end{figure*} 
In order to gain further insights into the dynamics
of the density field, we performed a scale-dependent analysis. 
Let $d\rho(r,x,t)=\rho(r+x,t)-\rho(r,t)$ be the density increment
at a scale $x$ at time $t$, and define the corresponding
structure functions as:
\begin{equation}
S_p(x,t)= <|d\rho(r,x,t)|^p>_r,\;\;\;p=1,2,..,5
\end{equation}
as well as scaling exponents $a_p$ via:
\begin{equation}
S_p(x,t) \sim S_p(x_0,t) (\frac{x}{x_0})^{a_p},\;\;\;1 \leq x_0<<L 
\end{equation}
where $x_0=int(d)$.
It is well known that a linear dependence $a_p=Kp$ indicates
a fractal process of order $K$ ($K=1$ for smooth, differentiable
processes, $K=1/2$ for standard diffusion), whereas
a non-linear dependence on $p$ would signal a multi-fractal
process instead \cite{diff}.
Our data show that $S_p(x,t) \sim x^{K(t)p}$, where $K(t)$ 
is a time-varying but scale-independent parameter.
This suggests that the density diffusion process is a 
fractal of dimension $K(t)$, with $K$ raising in the course of the evolution
from $0.08$ to the steady value $0.15$.
This also hints at a hierarchical organization
of density peaks and dips, which emerges spontaneously from
the non-linear coupling between the fluid motion
and the obstacles dynamics.

The fractal character is more neatly evidenced by
using extended self-similarity \cite{ess}. 
Indeed, in fig.~\ref{fig_fratt} we show the plots
of $S_p(x,t)$ vs $S_5(x,t)$ at two times for $p=1,2,..,5$. 
Data points fall on the straight lines of slope $5/p$, as 
expected in the case of fractal behavior.
The low values of $K(t)$ indicate that the density redistribution
is a {\it sub}-diffusive process, i.e. it proceeds more slowly
than Brownian motion \cite{diff}, as it is expected for motion
in a heterogeneous, disordered background.

\section{Conclusions}
In summary, we have presented a mesoscopic model of
fluid motion with random dynamical constraints.
Fluid motion is based on a lattice Boltzmann model, whereas
dynamical constraints are implemented via a control field
acting as a penetrable barrier for particle motion along
the links.
Despite its simplicity, our model displays some
features of disordered fluids, namely:
i) Onset of non-zero order parameter (density contrast) with
very small resilience to non-zero obstacle concentration
$c$ and impermeability $1-p_t$;
ii) Non-exponential, non-stationary time relaxation of
density correlation functions;
iii) Long-range spatial order, with non-Gaussian PDF's of
the density distribution,
iv) Sub-diffusive dynamics of the density distribution.

Although no strict correspondence with any specific physical system
can be claimed at this stage, the natural physical target of our model
are fluids in porous media with a dynamic morphology, nonlinearly
coupled to fluid motion.
For the future, we plan to investigate the behavior of the
present model for two and three-dimensional fluids.
In the long-term, we would also like to model glassy materials, although
this is certainly going to require new substantial
extensions of the simple model presented in this work.

\begin{acknowledgement}
Illuminating discussions with K. Binder, 
E. Marinari, G. Parisi and F. Sciortino are kindly acknowledged.
\end{acknowledgement}

\end{document}